\documentclass[english,12pt]{article}
\usepackage{color}
\usepackage{amssymb}
\usepackage{amsmath}
\usepackage{babel}
\usepackage{pifont}
\usepackage
[dvips,
colorlinks=true,
linkcolor=webgreen,
filecolor=webbrown,
citecolor=webgreen,
bookmarksopen=false,
]{hyperref}
\definecolor{webgreen}{rgb}{0,.5,0}
\definecolor{webbrown}{rgb}{.6,0,0}
\usepackage[T1]{fontenc}
\usepackage[cp1250]{inputenc}
\usepackage{array}
\usepackage{amssymb}
\date{}
\definecolor{arcolor}{cmyk}{0.05,0.95,0.9,0.1}
\title{Arbitrage risk induced by transaction costs}
\author{Edward W. Piotrowski\\ Institute of Theoretical Physics,
University of Bia\l ystok,\\ Lipowa 41, Pl 15424 Bia\l ystok,
Poland\\ e-mail: \href{mailto:ep@alpha.uwb.edu.pl}{ep@alpha.uwb.edu.pl}\\
 Jan S\l adkowski\\ Institute of Physics, University of Silesia, \\ Uniwersytecka
4, Pl 40007 Katowice, Poland \\ e-mail:
\href{mailto:sladk@us.edu.pl}{sladk@us.edu.pl} }
\usepackage[dvips]{graphicx}
\usepackage{pstricks}
\usepackage{pst-node}
\begin{document}
\maketitle
\def\Z{{\bf Z\!\!Z}}
\def\R{{\bf I\!R}}
\def\N{{\bf I\!N}}
\def\C{{\bf I\!\!\!\! C}}
\def\meter{\mbox{$\frown\hspace{-.9em}{\lower-.4ex\hbox{$_\nearrow$}}$}}
\begin{abstract}
We discuss the time evolution of quotation of stocks and
commodities and show that they form an Ising chain. We show that
transaction costs induce arbitrage risk that usually is neglected.
The full analysis of the portfolio theory is computationally
complex but the latest development in quantum computation theory
suggests that such a task can be performed on quantum computers.
\end{abstract}


 \vspace{5mm}

\section{Introduction}
One can simply define arbitrage as an opportunity of making profit
without any risk \cite{1}. But this definition has one flaw: it
neglects  transaction costs. And any market activity involves
costs (e.g\mbox{.} brokerage, taxes and others depending on the
established rules). Therefore there is always some uncertainty and
an arbitrageur cannot avoid risk. Below we will describe an
extremely profitable manipulation of a one asset market that
certainly fit this definition and show how brokerage can induce
risk. The method allows to make maximal profits in a fixed
interval $[0,k]$ (short-selling allows to make profits with
arbitrary price changes). We will analyze the associated risk by
introducing {\it canonical arbitrage portfolios} that admit Ising
model like description. Investigation of such models is difficult
from the computational point of view (the complication grows
exponentially in $k$)  but the latest development in quantum
computation seems to pave the way for finding effective methods of
solving the involved computational problems \cite{2}.

\section{Profit from brokerage-free transactions}
Standard descriptions of price movements, following Bachelier, use
the formalism of diffusion theory and random variables. Such an
approach to the problem involves the assumption of constancy of
the parameters of model during the interval used for their
estimation. Besides the use of the dispersion of the drifting
logarithm of price as a measure of the risk might be questioned. A
clairvoyant that knows the future evolution of prices would make
profits from any price movement and it would be difficult to
attribute any risk to her market activity. Rather, the level of
erroneousness of our decisions concerning the portfolio structure
should be used for that aim. Having this in mind we have proposed
a dual formulation of the portfolio theory and market prices
\cite{3}-\cite{6}. In this approach movements in prices are
regarded as deterministic according to their historical record and
the stochastic properties are attributed to portfolios. This
enables us to use the formalism of information theory and
thermodynamics. Due to the convenience of this approach we will
adopt it in the current paper. \\

Consider a game against the Rest of the World (that is the whole
market) that consist in alternate buying and selling of the same
commodity. Let $h_m\negthinspace:=\ln\frac{c_m}{c_{m-1}}$ denote
logarithms of the prices dictated by the market of this commodity
at successive moments
$m\negthinspace=\negthinspace1,2,\ldots,\negthinspace k$\/\,. If the costs
of transactions are zero (or negligible) then the player's profit
(actually a loss because for future convenience we will fix the
sign in $(\ref{forha})$ according to the standard physical
convention) in the interval $[0,k]$ is given by
\begin{equation}
H(n_1,\ldots,n_k):=-\sum\limits_{m=1}^k h_m n_m\,. \label{forha}
\end{equation}
The elements of the sequence $(n_m)$ take the value $0$ or $1$
if the player possesses  money or the commodity at the moment $m$,
respectively. The sequence $(n_m)$ defines the player's strategy
in a unique way and any $(n_1,\ldots,n_k)$ describes a pure
strategy. Generalization to a more realistic situation where more
commodities are available is trivial but besides complication of
formulas is irrelevant to the conclusion and will not be
considered here.

\section{The thermodynamics of portfolios}
Any mixed strategy can be parameterized in a unique way by $2^k$
weights $p_{n_1,\ldots,n_k}$ giving the contributions of pure
strategies. Let us consider as equivalent all strategies that for
a given price sequence $(h_1,\ldots,h_k\negthinspace)$ bring the
same profit. We will call the equivalence classes of
portfolios defined in this way
{\it the canonical portfolios}\/. Any canonical portfolio
has maximal information entropy
\begin{equation}
S_{(p_{n_1,\ldots,n_k}\negthinspace)}:=-E(\ln
p_{n_1,\ldots,n_k\negthinspace})\, \label{deffendr}
\end{equation}
and can in a sense be regarded as an equilibrium state for
portfolios in its class (the player rejects any superfluous from
the market point of view) information. Claude Shannon's entropy
$S_{(p_{n_1,\ldots,n_k})}$ rooted in cryptographic information
theory is proportional to the minimal length of the compressed by
the greedy Huffman algorithm computer code that contains
information about the portfolio \cite{7}. Therefore it seems to be
reasonable to accept the risk incurred of investing in a given
canonical portfolio as a measure of risk for the whole class it
represents. The explicit form of a canonical portfolio can be
found by the Lagrange multipliers method that leads to the
requirement of vanishing of the following differential form:
$$
d S_{(p_{n_1,\ldots,n_k})}-\beta\, dE(H(n_1,\ldots,n_k))-\,\zeta\,
d
\negthinspace\negthinspace\negthinspace\negthinspace\negthinspace
\sum\limits_{\phantom{a}n_1,\ldots,n_k=0}^{1}
\negthinspace\negthinspace\negthinspace\negthinspace
p_{n_1,\ldots,n_k}=
$$\nopagebreak
$$
-\negthinspace\negthinspace\sum\limits_{\phantom{a}n_1,\ldots,n_k=0}^{1}
\negthinspace\negthinspace\bigl(\,\ln p_{n_1,\ldots,n_k} +1
+\beta\,H(n_1,\ldots,n_k) +\zeta\bigr)\,dp_{n_1,\ldots,n_k}=0\,,
$$
where $\beta$  and $\zeta$  are Lagrange multipliers. It follows
that the sum  $\ln p_{n_1,\ldots,n_k} +1 +\beta\,H(n_1,\ldots,n_k)
+\zeta$ should  vanish independently of the values of
$dp_{n_1,\ldots,n_k}$. Therefore the equation
$$\ln p_{n_1,\ldots,n_k}\negthinspace +1
+\beta\,H(n_1,\ldots,n_k) +\zeta=0
$$
allows to find the dependence of the weights $p_{n_1,\ldots,n_k}$
on the profits $-H(n_1,\ldots,n_k)$ resulting from pure
strategies:
$$
p_{n_1,\ldots,n_k}={\mathrm
e}^{-\beta\,H(n_1,\ldots,n_k)-\zeta-1}.
$$
The Lagrange multiplier  $\zeta$ can be eliminated by
normalization of the weights. This leads to the Gibbs distribution
function \cite{8}:
$$ p_{n_1,\ldots,n_k}=\frac{{\mathrm
e}^{-\beta\,H(n_1,\ldots,n_k)}}{
\sum\limits_{\phantom{a}n_1,\ldots,n_k=0}^{1}
\negthinspace\negthinspace {\mathrm
e}^{-\beta\,H(n_1,\ldots,n_k)}}\,.
$$
Note that we have put no restriction on the properties of
$H(n_1,\ldots,n_k)$. The complete information about this random
variable is contained in the statistical sum
$$
Z(\beta):=\sum\limits_{\phantom{a}n_1,\ldots,n_k=0}^{1}
\negthinspace\negthinspace\text{e}^{-\beta\, H(n_1,\ldots,n_k)}\,,
$$
because its logarithm is the cumulant-generating function (the
moments of  $H(n_1,\ldots,n_k)$ are given by  $(-1)^n\,\frac{d^n\ln
Z}{d\beta^n}\,$). Physicists used to call the inverse of the
Lagrange multiplier  $\beta$ the temperature $T$. The expectation
of the profit $-E(H(n_1,\ldots,n_k))$ is a decreasing function of
$T$ and in the limit $T\negthinspace\rightarrow\negthinspace0^+$
it reaches its maximum. It is easy to notice that the statistical
sum $Z(\beta)$ factorizes for the profit function given by
$(\ref{forha})$:
$$
\sum\limits_{\phantom{a}n_1,\ldots,n_k=0}^{1}
\negthinspace\negthinspace\text{e}^{
\,\beta\negthinspace \sum\limits_{m=1}^k h_m
n_m}= \sum\limits_{\phantom{a}\hspace{-1.2em}n_1,\ldots,n_k=0}^{1}
\negthinspace\prod\limits_{\phantom{a}m=1}^k\text{e}^{\,\beta\, h_m
n_m}= \prod\limits_{m=1}^k \sum\limits_{n_m=0}^1\text{e}^{\,\beta\,
h_m n_m}\,.
$$
This means that the profits made at different moments are
independent and there exist a risk-free pure arbitrage strategy of
the form: {\em keep the commodity only if the prices are
increasing} (that is $n_m\negthinspace=\negthinspace
\frac{1\,+\,\text{sign}\,h_m}{2}$\,). The opposing strategy
$(n_m\negthinspace=\negthinspace
\frac{1\,-\,\text{sign}\,h_m}{2}\,)$ defining a canonical
portfolio for $T\negthinspace\rightarrow\negthinspace0^-$ forms a
risk-free strategy for short positions. The canonical portfolio
representing monkey strategies has infinite temperature,
$T\negthinspace=\negthinspace\pm\infty$. To translate the
Markowitz portfolio theory into our thermodynamical language we
should include in an explicit way   the portfolio risk measured by
its second cumulant moment $\bigl(H\negthinspace-\negthinspace
E(H)\bigr)^2$ (and the corresponding Lagrange multiplier!) besides
the random variable $H(n_1,\ldots,n_k)$. For our aims it suffices
to consider only the equilibrium variant of canonical portfolios
with average\footnote{Markowitz theory deals with effective
portfolios that are characterized by minimal risk at a given
profit level.} risk given by $\frac{\partial^2\ln
Z}{\partial\beta^2}$\,.
\section{Non-zero transactions costs}
If we take transaction costs into consideration\footnote{For
simplicity we consider only the case of cost constant per unit of
the commodity.}  then Eq.~$(\ref{forha})$ should be replaced by
the more general formula:
\begin{equation}
-\sum\limits_{m=1}^k h_m n_m-j\,(n_{m-1}\negthinspace\oplus
n_m\negthinspace)\,\,\, \rightarrow\,\,
H(n_1,\ldots,n_k\negthinspace)\,, \label{forha1}
\end{equation}
where $n_0\negthinspace:=\negthinspace n_k$ for periodic boundary
conditions (otherwise $n_0$ should be fixed arbitrary). $\oplus$
denotes addition modulo 2, and the constant
$j\negthinspace>\negthinspace0$ is equal to the logarithm of the
cost of a single transaction. The careful reader will certainly
notice that Eq.~$(\ref{forha1})$ represent an hamiltonian of an
Ising chain \cite{9} (the shift in the sequence $(n_m)$ by
$-\frac{1}{2}$ introduces only an unimportant constant to the
formula). Now the statistical sum $Z(\beta)$ cannot be factorized
in terms of contributions from separate moments. Instead we should
use transition matrices that depend on the immediate moments
$m\negthinspace-\negthinspace1$ i $m$\,. Then for a convenient
periodic boundary conditions we arrive at the formula that
expresses $Z(\beta)$ as a trace of the product of transition
matrices:$$
 Z(\beta)=\negthinspace\negthinspace\negthinspace\sum
 \limits_{\phantom{a}n_1,\ldots,n_k=0}^{1} \negthinspace\negthinspace
 M(1)_{n_k\,n_1}M(2)_{n_1\,n_2}\cdots
 M(k)_{n_{k-1}\,n_k}\,=\, \text{Tr}\prod\limits_{m=1}^k M(m)\,,
$$
where
$$
M(m)_{n_{m-1}\,n_{m}}:=\text{e}^{\beta(h_m\, n_m-j\,
(n_{m-1}\oplus\, n_m) )}\,.
$$
Unfortunately, the entries of the matrices $M(m)$ depend on time
via $h_m$ and the analysis of the proper value problem does not
lead to any compact form of the statistical sum (except for the
uninteresting case of the  constant sequence  $(h_m)$).
\section{The (min,+) algebra of portfolios}
The definition of entropy $(\ref{deffendr})$ allows to find the
following relation among the entropy, average profit and the
statistical sum:
\begin{equation}
E(H)+T\,\ln Z = T\,S\,. \label{tyttoly}
\end{equation}
The entropy is positive and bounded from above ($S\leq k\ln 2$)
therefore Eq.~$(\ref{tyttoly})$ can be used to determine the
strategies giving maximal profits:
$$
H_\pm:=\lim\limits_{T\rightarrow0^{\pm}}E(H)=-\lim\limits_{T\rightarrow0^{\pm}}T\,\ln
Z=
\lim\limits_{\beta\rightarrow\pm\infty}\log_{\text{e}^{-\beta}}Z\,.
$$
Elementary properties of logarithms\footnote{For short positions
strategies ($T\rightarrow0^-$) we should find the limit
$\lim\limits_{\varepsilon\rightarrow
\infty}\log_\varepsilon(\varepsilon^a\negthinspace+\varepsilon^b)=
\max(a,b)$}
$$
\log_\varepsilon(\varepsilon^a\,\varepsilon^b)=a+b\,,\,\,\,\,\,
\lim\limits_{\varepsilon\rightarrow0^+}\log_\varepsilon(\varepsilon^a+\varepsilon^b)=
\min(a,b)
$$
imply that the full information about the most profitable strategy
is given by the product of logarithms of transition matrices
$$
\tilde{M}(m)_{n_{m-1}\,n_{m}}\negthinspace:=
\log_{\text{e}^{-\beta}} M(m)_{n_{m-1}\,n_{m}}=-h_m\, n_m+j\,
(n_{m-1} \negthinspace\oplus n_m )\,
$$ if we replace addition of real numbers by the operation
min of taking the minimal element of them and  multiplication of
numbers  by their sum  (that is by using the (min,+) algebra
\cite{10})
\begin{equation}
\bigl(\tilde{M}(m)\times\negthinspace\,\tilde{M}(m\negthinspace+
\negthinspace1)\bigr)_{n_{m-1}\,n_{m+1}} :=\min\limits_{n_m}\bigl(
\tilde{M}(m)_{n_{m-1}\,n_{m}}\negthinspace+\tilde{M}(m\negthinspace+\negthinspace
1)_{n_{m}\,n_{m+1}}\bigr)\,. \label{opperra}
\end{equation}
Investigation of the matrix elements contributing to the "product"
allows to reconstruct the sequence $(n_1,\ldots,n_k)$ and its
minimal element will correspond to the maximal available profit in
the game.

\section{ Arbitrage risk}
For a given price  $(h_1,\ldots,\negthinspace h_k)$ let us call
{\em the potential arbitrage strategy}\/ any strategy that if
completed with element corresponding to moments
$k'\negthinspace>\negthinspace k$  might turn out to be the
strategy giving maximal profit for the hamiltonian
$H(n_1,\ldots,n_k,\ldots,\negthinspace n_{k'}\negthinspace)$. In
our case there are only four potential arbitrage strategies if the
initial value $n_0$  is not fixed. In general, in a market with
$N\negthinspace-\negthinspace1$ commodities there are $2N$ such
strategies. It is easy to notice that for a given price sequence
$(h_1,\ldots,h_k)$ potential arbitrage strategy has the form
$$
(0,1,1,0,1,0,0,n_{k-l+1},n_{k-l+2},\ldots,n_k)\,,
$$
and can be decomposed into two parts. The first one constructed
according to the  knowledge of the sequence
$(h_1,\ldots,\negthinspace h_k)$ and the second one of the length
$l$. We will call $l$ the {\em coherence depth}\/. The functional
dependence of the coherence depth $l$ on the costs $j$ might form
an interesting market indicator of structure of price movements.
The final sequence $(n_{k-l+2},\ldots,n_k)$ can be determined only
if prices $h_m$ are known for $m\negthinspace >\negthinspace k$.
Therefore any potential arbitrage strategy is an optimal strategy
for a player whose profits are known only up to the moment $k$. In
that sense non-vanishing transaction costs involve arbitrage risk
that might be caused, for example, by the finite maturity time of
contracts or splitting of orders. \\
The algorithm for finding potential arbitrage strategies, the
respective profits and coherence depths is fast because uses only
addition of matrices ("product") which is linear in $k$. But this
is insufficient for the risk and profits analysis of all
portfolios equivalent to potential arbitrage portfolios. To
analyze the arbitrage opportunities we should consider the whole
low temperature sector of canonical portfolios
$T\negthinspace\in\negthinspace(0,T_+]$, where $E_{T_+}(H):=
\max_{n_0,\,n_k}\bigl(\tilde{M}(1)\times\cdots\times\negthinspace\tilde{M}(k)\bigr)_{n_{0}\,n_{k}}$
because it is the minimal set that contains all potential
arbitrage portfolios. Unfortunately due to the lack of compact
form of the statistical sum, the knowledge of canonical arbitrage
portfolios requires performing of $3^k$ arithmetical operations
what is a difficult computational task.
\section{Simulations of canonical portfolios}
Simulations are usually  perceived as  modelling of real processes
by a Turing machines. But complexity of various phenomena shows
the limits of effective polynomial algorithms. Is seems that the
future will reverse the roles: we will compute by simulations
perceived as measurements of appropriate (physical?) phenomena. In
fact  such methods have been used for centuries\footnote{Gaudy
constructed the vault of the Sagrada Familia Church in Barcelona
with the help of gravity.}. The model discussed  above can be
easily associated with quantum computation (and games).
Calculations for portfolios should take into consideration all
available  pure strategies whose number grows exponentially in $k$
($2^k$). Therefore the classical Turing machines are of little
use. One of the future possibilities might be exploration of
nano-structures having properties of Ising chains. Changes of
local magnetic fields $h_m$ and controlling temperature may allow
for effective determination of profits and strategies for players
of various abilities (measured by their temperature \cite{3,4}).
The values of the parameters $n_m$ would be found  by measurements
of magnetic moments. Another effective method might consist in
using quantum parallelism for simultaneous determination of all
$2^k$ components of the statistical sum. Quantum computation would
use superpositions of  $k$ qubit quantum states
$$
\mathbb{C}P^{2^k\negthinspace-1}\ni|\psi\rangle:=
\negthinspace\negthinspace\negthinspace\sum
\limits_{n_1,\ldots ,
n_k=0}^{1}\negthinspace\negthinspace\negthinspace
\text{e}^{\,\text{i}\,\varphi_{n_1\ldots n_k}+\tfrac{\beta}{2}\sum\limits_{m=1}^k
(h_m n_m-j\,(n_{m-1}\negthinspace\oplus
n_m\negthinspace))}|n_1\rangle\otimes\ldots\otimes |n_k\rangle
$$
with arbitrary phases  $\varphi_{n_1\ldots n_k}$. Measurements of
the states $|\psi\rangle$ would allow to identify for a given
portfolio all important leading terms in the statistical sum. The
paper \cite{11} presents  analysis of the problem of simulation of
an Ising chain on a quantum computer. One can easily identify the
unitary transformations used there  with transition matrices for
probability amplitudes. Details of such computations and their
interpretation in term of quantum market games will be presented
in a separate paper (cf \cite{12}).


\end{document}